\documentclass[useAMS,usenatbib]{mn2e}

\usepackage{graphicx}
\usepackage{epstopdf}
\usepackage{color}
\usepackage[pdftex, hidelinks, draft]{hyperref}

\title[Unbiased FIR observations of GRB hosts]{Far-infrared observations of an unbiased sample of gamma-ray burst host galaxies}
\author[S. A. Kohn et al.]{S.~A. Kohn,$^{1,2}$\thanks{E-mail: saulkohn@sas.upenn.edu} M.~J. Micha{\l}owski,$^{1}$ N. Bourne,$^{1}$ M. Baes,$^{3}$ J. Fritz,$^{3,4}$ A. Cooray,$^{5}$ 
\newauthor I. de Looze,$^{3,6}$ G. De Zotti,$^{7,8}$ H. Dannerbauer,$^{9}$ L. Dunne,$^{1,10}$ S. Dye,$^{11}$ S. Eales,$^{12}$ 
\newauthor C. Furlanetto,$^{11}$ J. Gonzalez-Nuevo,$^{13,7}$ E. Ibar,$^{14}$ R.~J. Ivison$^{15,1}$ S.~J. Maddox,$^{1,10}$ 
\newauthor D. Scott,$^{16}$  D.~J.~B. Smith,$^{17}$ M.~W.~L. Smith,$^{12}$ M. Symeonidis,$^{18, 19}$ E. Valiante$^{12}$\\
$^{1}$Institute for Astronomy, University of Edinburgh, Royal Observatory, Blackford Hill, Edinburgh EH9 3HJ, UK\\
$^{2}$Department of Physics and Astronomy, University of Pennsylvania, Philadelphia, PA, 19104, USA\\
$^{3}$Sterrenkundig Observatorium, Universiteit Gent, Krijgslaan 281 S9, B-9000 Gent, Belgium\\
$^{4}$Centro de Radioastronom\'\i a y Astrof\'\i sica, CRyA, UNAM, Campus Morelia, A.P. 3-72, C.P. 58089, Michoac\'an, Mexico\\
$^{5}$Center for Cosmology, Department of Physics and Astronomy, University of California, Irvine, CA 92697, USA\\
$^{6}$Institute of Astronomy, University of Cambridge, Madingley Road, Cambridge, CB3 0HA, UK\\
$^{7}$SISSA, Via Bonomea 265, I-34136, Trieste, Italy\\
$^{8}$INAF-Osservatorio Astronomico di Padova, Vicolo dell'Osservatorio 5, I-35122 Padova, Italy\\
$^{9}$Institut f{\"u}r Astrophysik, Universit{\"a}t Wien, T{\"u}rkenschanzstra{\ss}e 17, 1180 Wien, Austria\\
$^{10}$University of Canterbury, Department of Physics and Astronomy, Private Bag 4800, Christchurch, 8041, New Zealand\\
$^{11}$School of Physics \& Astronomy, University of Nottingham, University Park, Nottingham NG7 2RD, UK\\
$^{12}$Cardiff School of Physics and Astronomy, Cardiff University, Queen's Buildings, The Parade, Cardiff, CF24 3AA, UK\\
$^{13}$Inst. de Fisica de Cantabria (CSIC-UC), Avda. los Castros s/n, 39005 Santander, Spain\\
$^{14}$Instituto de F\'isica y Astronom\'ia, Universidad de Valpara\'iso, Avda. Gran Breta\~na 1111, Valpara\'iso, 5030, Chile\\
$^{15}$European Southern Observatory, Karl Schwarzschild Strasse 2, D-85748 Garching, Germany\\
$^{16}$Department of Physics \& Astronomy, University of British Columbia, Vancouver, BC, V6T 1Z1 Canada\\
$^{17}$Centre for Astrophysics, Science \& Technology Research Institute, University of Hertfordshire, Hatfield, Herts, AL10 9AB, UK\\
$^{18}$Astronomy Centre, Department of Physics \& Astronomy, University of Sussex, Brighton BN1 9QH, UK\\
$^{19}$Mullard Space Science Laboratory, University College London, Holmbury St. Mary, Dorking, Surrey RH5 6NT, UK\\
}

\begin{document}

\date{}

\pagerange{\pageref{firstpage}--\pageref{lastpage}} \pubyear{2014}
\maketitle

\label{firstpage}

\begin{abstract}
Gamma-ray bursts (GRBs) are the most energetic phenomena in the Universe; believed to result from the collapse and subsequent explosion of massive stars. 
Even though it has profound consequences for our understanding of their nature and selection biases, little is known about the dust properties of the galaxies hosting GRBs. 
We present analysis of the far-infrared properties of an unbiased sample of 20 \textit{BeppoSAX} and \textit{Swift} GRB host galaxies (at an average redshift of $z\,=\,3.1$) located in the {\it Herschel} Astrophysical Terahertz Large Area Survey, the {\it Herschel}  Virgo Cluster Survey, the {\it Herschel}  Fornax Cluster Survey, the {\it Herschel}  Stripe 82 Survey and the {\it Herschel}  Multi-tiered Extragalactic Survey, totalling $880$ deg$^2$, or $\sim 3$\% of the sky in total.  Our sample selection is serendipitous, based only on whether the X-ray position of a GRB lies within a large-scale {\it Herschel} survey -- therefore our sample can be considered completely unbiased.
Using deep data at wavelengths of 100\,--\,500$\,\mu$m, we tentatively detected 1 out of 20 GRB hosts located in these fields. 
We constrain their dust masses and star formation rates (SFRs), and discuss these in the context of recent measurements of submillimetre galaxies and ultraluminous infrared galaxies. The average far-infrared flux of our sample gives an upper limit on SFR of $<114$ M$_{\sun}\,$yr$^{-1}$. The detection rate of GRB hosts is consistent with that predicted assuming that GRBs trace the cosmic SFR density in an unbiased way, i.e. that the fraction of GRB hosts with $\mbox{SFR}>500\,{\rm M}\odot\,\mbox{yr}^{-1}$ is consistent with the contribution of such luminous galaxies to the cosmic star formation density.
\end{abstract}


\section{Introduction}\label{intro}

Long-duration gamma-ray bursts (GRBs) have, in many cases, been proven to be connected with supernovae \citep[e.g.][]{Galama.98, Hjorth.03}, suggesting that their progenitor stars are short-lived \citep{Heger.03}. As a result, GRBs are expected to be located in regions undergoing active star formation, making GRBs potentially ideal tracers of star formation across a large range of redshifts and helping us to chart the star-forming history of our Universe \citep[e.g.][]{Wijers.98, Yonetoku.04}.

Many properties of GRBs are useful for accomplishing the above task. Their extremely high luminosity at gamma-ray wavelengths allows for their detection out to very high redshift, with minimal extinction from gas or dust. The average GRB redshift is $z\sim2.2$ \citep{Fynbo.09, Jakobsson.06, Jakobsson.12, Hjorth.12}, but in principle GRBs should be visible out to $z\geq15$--$20$ \citep{Lamb.00}, and in practice they have been observed out to $z\sim9$ \citep{Tanvir.09, Salvaterra.09, Cucchiara.11}. Moreover, detections of GRBs are independent of the luminosity of their host galaxies, which allows us to probe fainter galaxies than in typical flux-limited samples \citep[e.g.][]{Tanvir.12}. Previous studies have found that GRB hosts are mostly faint and blue \citep{leFloch.03}, with the GRB occurring in (rest-frame) UV-bright regions \citep{Bloom.02, Fruchter.06, Leloudas.10, Leloudas.11}, which is consistent with active star formation. More recent studies \citep[e.g.][]{Greiner.11, Kruhler.11, Rossi.12, Perley.13} show that GRB hosts span a wider range of properties, suggesting that previous optically-selected host surveys were biased towards these faint blue hosts.

Before we can use GRBs to quantitatively trace star formation across cosmic history, such biases of GRB and GRB host galaxy samples must be well understood. Using The Optically Unbiased GRB Host survey \citep[TOUGH;][]{Hjorth.12}, \cite{Michalowski.12} and \cite{Perley.14} analysed the radio-derived star formation rates (SFRs) of GRB hosts at $z<2$. They found that the distributions of SFRs of GRB hosts were consistent with that of other star-forming galaxies at $z<2$, suggesting that GRBs may be able to trace cosmic star formation; however they made clear that further study of potential biases in the morphology and metallicity of GRB hosts is required. Similarly, \cite{Hunt.14} and \cite{Schady.14} concluded that SFRs in GRB hosts are consistent with other star-forming galaxies, and found no strong evidence for GRB hosts being biased tracers of the global star formation rate density (SFRD). However, their sample favoured infrared-bright hosts. In their study of the hosts of dust-obscured GRBs, \cite{Perley.13, Perley.14} found that they are not massive enough to bring the overall GRB host population into the expected territory of an unbiased SFR-tracing population for $z\sim1$. They found that GRB hosts appear to be biased towards low-mass galaxies and suggested that the GRB rate relative to SFR is highly dependent on host-galaxy environment at this redshift, and highlighted that further studies of metallicity are needed. These findings are corroborated by \cite{Boissier.13}.

Here we attempt to study the distribution of SFRs of GRB hosts using the unbiased sample of GRBs that are located inside wide-area {\it Herschel}\footnote{{\it Herschel} is an \textit{ESA space observatory} with science instruments provided by European-led Principal Investigator consortia and with important participation from NASA.} \citep{Pilbratt.10} surveys. It is important to study this aspect at both ultraviolet/optical and infrared wavelength, since these give access to unobscured and dust-obscured star formation activity, respectively. The latter is still poorly understood for GRB hosts, where deep observations and detections have been largely limited to low redshifts and biased samples \citep{Frail.02, Berger.03, Tanvir.04, deUgarte.12, Wang.12, H.14, Hunt.14, Michalowski.14, Schady.14, Sym.14}. The objectives of this paper are to: (1) measure the dust-obscured SFRs and dust masses of GRB host galaxies using an unbiased sample; and (2) test whether or not GRBs are unbiased tracers of cosmic star formation at $z>2$.

The layout of this paper is as follows. Section~\ref{samp} contains an overview of the \textit{Herschel} surveys whose data we used to acquire our sample. We then discuss the measured properties of the GRB hosts in Section~\ref{res}, before moving on to a more extensive discussion of these properties as compared to recent analyses of other galaxy types in Section~\ref{disc}. We provide out conclusions in Section~\ref{conc}.

We use a cosmological model of $H_{0}\,$=$\,70\,\mbox{km\,s}^{-1}\mbox{Mpc}^{-1}$, $\Omega_{\Lambda}=0.7$ and $\Omega_{\rm m}=0.3$, and a Salpeter \citep{Salpeter.55} initial mass function.\\

\begin{table}
\begin{footnotesize}

\caption{Details of {\it Herschel} surveys used}
\label{tab:surveys}
\begin{tabular}{llllllll}
\hline
Survey	&	Area	&	\multicolumn{5}{c}{Total noise (mJy beam$^{-1}$)$^a$}&	Ref. \\
\cline{3-7}
		&	(deg$^2$)&	100 &	160 & 250 & 350 & 500 & \\
\hline
H-ATLAS & 600 & 25 &30 & 7.2 & 8.1 & 8.8 & 1, 2 \\
HeFoCS & 16 & 9.9 & 9.2 & 8.9 & 9.4 & 10.2 & 3\\
HerMES$^{b}$ & 100 &-- &-- &6.4 &6.8 &7.6 & 4 \\
HerS$^{c}$ & 79 &-- &-- &10.7 &10.3 &12.3 & 5\\
HeViCS & 84 & 23 & 13 & 6.6 & 7.3 & 8.0 & 6 \\
\hline
\end{tabular}
\\
$^a$ The wavelength in microns of each band is given in the header.\\
$^{b}$ HerMES data have different noise levels depending upon the field surveyed. The average noise levels for each band are presented here, given in reference (4).\\
$^{c}$ Coverage of the HerS maps is not uniform. The average noise levels are presented here for the deeper part of the survey (where GRB 060908 is located, see Table~\ref{table_sample} and reference 5).\\
Hyphens indicate that data were not taken or not available in that band.\\ 
\textbf{References:} (1)~\cite{Ibar.10}; (2)~\cite{Pascale.11}; (3)~\cite{Fuller.14}; (4)~\cite{Smith.12}; (5)~\cite{Viero.14}, and (6)~\cite{Auld.13}.

\end{footnotesize}
\end{table}

\section{Sample and Data}\label{samp}

\begin{table*}																			
  \caption{GRB sample}																			
  \label{table_sample}																			
  \begin{tabular}{lllllll}																			
        \hline																			
	GRB	&	$\alpha$	 &	$\delta$	 &	90\% error	&	{\it Herschel}	&	$z$		&	Reference					\\
		&	(J2000)	&	(J2000)	&	(arcsec)	&	Field	&		&							\\
	\hline																		
	990308	&	185.7976667	&	6.7347500	&	0.3 &	HeViCS	&		&	\citealt{Schaefer.99}				\\
	050412	&	181.1053333	&	$-$1.2001667	&	3.7	&	GAMA12	&		&	\citealt{Hjorth.12}				\\
	050522	&	200.1440417	&	24.7890833	&	3.8	&	NGP	&		&	\citealt{Butler.07}				\\
	051001	&	350.9530417	&	$-$31.5231389	&	1.5	&	SGP	&	2.4296	&	\citealt{Butler.07, Hjorth.12}		\\
	060206	&	202.9312917	&	35.0507778	&	0.8	&	NGP	&	4.048	&	\citealt{Butler.07, Fynbo.09}		\\
	060908	&	31.8267500	&	0.34227778	&	1.4	&	HerS	&	1.884	&	\citealt{Hjorth.12, Jakobsson.12}		\\
	070611	&	1.99241667	&	$-$29.7554722	&	1.8	&	SGP	&	2.0394	&	\citealt{Butler.07, Hjorth.12}		\\
	070810A	&	189.9635000 &	10.7508889	&	1.4	&	HeViCS	& 2.17		&	\citealt{Hjorth.12, Schady.12}				\\
	070911	&	25.8095417	&	$-$33.4841389	&	3.1	&	SGP	&		&	\citealt{Butler.07}				\\
	071028B	&	354.1617917	&	$-$31.6203611	&	0.3	&	SGP	&		&	\citealt{Clemens.11}				\\
	080310	&	220.0575000 &	$-$0.1748889	&	0.6	&	GAMA15	&	2.42	&	\citealt{Littlejohns.12,	Fox.08}		\\
	091130B	&	203.1481250	&	34.0885278	&	0.6	&	NGP	&		&	\citealt{Butler.07}				\\
	110128A	&	193.8962917	&	28.0654444	&	0.4	&	NGP	&	2.339	&	\citealt{Butler.07, Sparre.11}		\\
	110407A	&	186.0313917	&	15.7118417	&	1.0	&	HeViCS	&		&	\citealt{Chuang.11}				\\
	111204A	&	336.6283750	&	$-$31.3748056	&	1.9	&	SGP	&		&	\citealt{Sonbas.11}				\\
	120703A	&	339.3567083	&	$-$29.7232500	&	0.5	&	SGP	&		&	\citealt{Xu.12}				\\
	120927A	&	136.6137500	&	0.4161944	&	1.4	&	GAMA09	&		&	\citealt{Beardmore.12}				\\
	121211A	&	195.5332917	&	30.1485000	&	0.5	&	NGP	&		&	\citealt{Chester.12}				\\
	130502A	&	138.5689583	&	$-$0.1233056	&	1.5	&	GAMA09	&		&	\citealt{Beardmore.13}				\\
	140102A	&	211.9193750	&	1.3332778	&	0.5	&	GAMA15	&		&	\citealt{Hagen.14}				\\
	140515A	&	186.0650000	&	15.1045556	&	1.8	&	HeViCS	&	6.327	&	\citealt{Butler.07, Chornock.14}		\\
        \hline																			
  \end{tabular}																																	
\end{table*}																			

The {\it Herschel} Astrophysical Terahertz Large Area Survey \citep[H-ATLAS;][]{Eales.10} is the largest open-time survey conducted by the {\it Herschel Space Observatory}. It covers approximately $600$ deg$^2$ in the far-infrared (FIR) and submillimetre wavelengths, using PACS ($100\,\mu$m, $160\,\mu$m; \citealt{Poglitsch.10}) and SPIRE ($250\,\mu$m, $350\,\mu$m and $500\,\mu$m; \citealt{Griffin.10}). The {\it Herschel} Virgo Cluster Survey \citep[HeViCS;][]{Davies.10} used PACS and SPIRE to survey a total area of $84$ deg$^2$ \citep{Auld.13, Baes.14} of the Virgo galaxy cluster, with diminishing sensitivities beyond the central $55$ deg$^2$. Similarly, the {\it Herschel} Fornax Cluster Survey \citep[HeFoCS;][]{Davies.13} used PACS and SPIRE to survey a total area of $16$ deg$^2$ in the Fornax cluster. The {\it Herschel} Stripe 82 Survey \citep[HerS;][]{Viero.14} covered $79$ deg$^2$ along the Sloan Digital Sky Survey \citep[SDSS;][]{York.00} `Stripe 82' field using SPIRE. The {\it Herschel} Multi-tiered Extragalactic Survey \citep[HerMES;][]{Oliver.12, Rosebloom.10, Smith.12, Viero.13, Wang.13} used SPIRE to observe 29 fields, covering $\sim100$ deg$^2$ in total. 

In this study, we use H-ATLAS GAMA, NGP and SGP data \citep[][Bourne et al. in prep, Valiante et al. in prep.]{Ibar.10, Pascale.11, Rigby.11, Smith.11}, all observed HeViCS fields (V1 to V4), the HeFoCS field \citep{Davies.13}, HerS maps\footnote{\url{www.astro.caltech.edu/hers/}} and the second major HerMES data release\footnote{\url{hedam.lam.fr/HerMES/}} \citep{Smith.12}. 
It is important to note that each of these surveys used different numbers of cross-scans with the instruments aboard \textit{Herschel}, resulting in different levels of sensitivity. These differences are summarized in Table~\ref{tab:surveys}.

\cite{Butler.07} presented a catalogue of refined X-ray positions of \textit{Swift}/XRT \citep{Gehrels.04, Burrows.05} GRBs\footnote{The catalogue (\url{butler.lab.asu.edu//Swift/xrt_pos.html}) is constantly updated. We accessed the catalogue on 2014 July 25. At this time, it had $780$ entries.}. Half of the 90\% confidence error region radii are $<2.2$ arcsec. It has been shown by \cite{Fynbo.09} and \cite{Hjorth.12} that such X-ray selection, as opposed to optical selection, does not introduce any bias, as long as close to 100\% of the gamma-ray triggered GRBs then have good X-ray localizations obtained for them. Indeed, nearly all GRBs are detected in X-ray when prompt observations are available.

We selected all GRBs with X-ray positions \citep{Butler.07} inside the H-ATLAS, HeViCS and HerS fields, which amounted to 21 GRBs. GRB 990308 occurred within the HeViCS field \citep{Schaefer.99}. As this is a pre-\textit{Swift} burst, in order to not corrupt the unbiased nature of our sample we simply provide our measurements of its flux density here, and it does not enter into any other calculations. No GRB positions overlapped with any HeFoCS or HerMES fields. The catalogue contains $780$ GRBs over the entire sky, so in $880\,\mbox{deg}^2$ we expect to find $780\times 880/41253\simeq 17$ GRBs, which is close to the number of $20$ that we found (i.e. excluding GRB 990308). Our sample selection process may be considered unbiased, since the only selection criterion applied was positional.  Table~\ref{table_sample} shows the positions and any known redshifts of the GRBs in our sample; Fig.~1 shows the {\it Herschel} data. For seven of our GRBs redshifts are known, and only for these cases do we provide physical properties. 

\begin{figure*}
\begin{center}
\includegraphics[width=0.8\textwidth,clip]{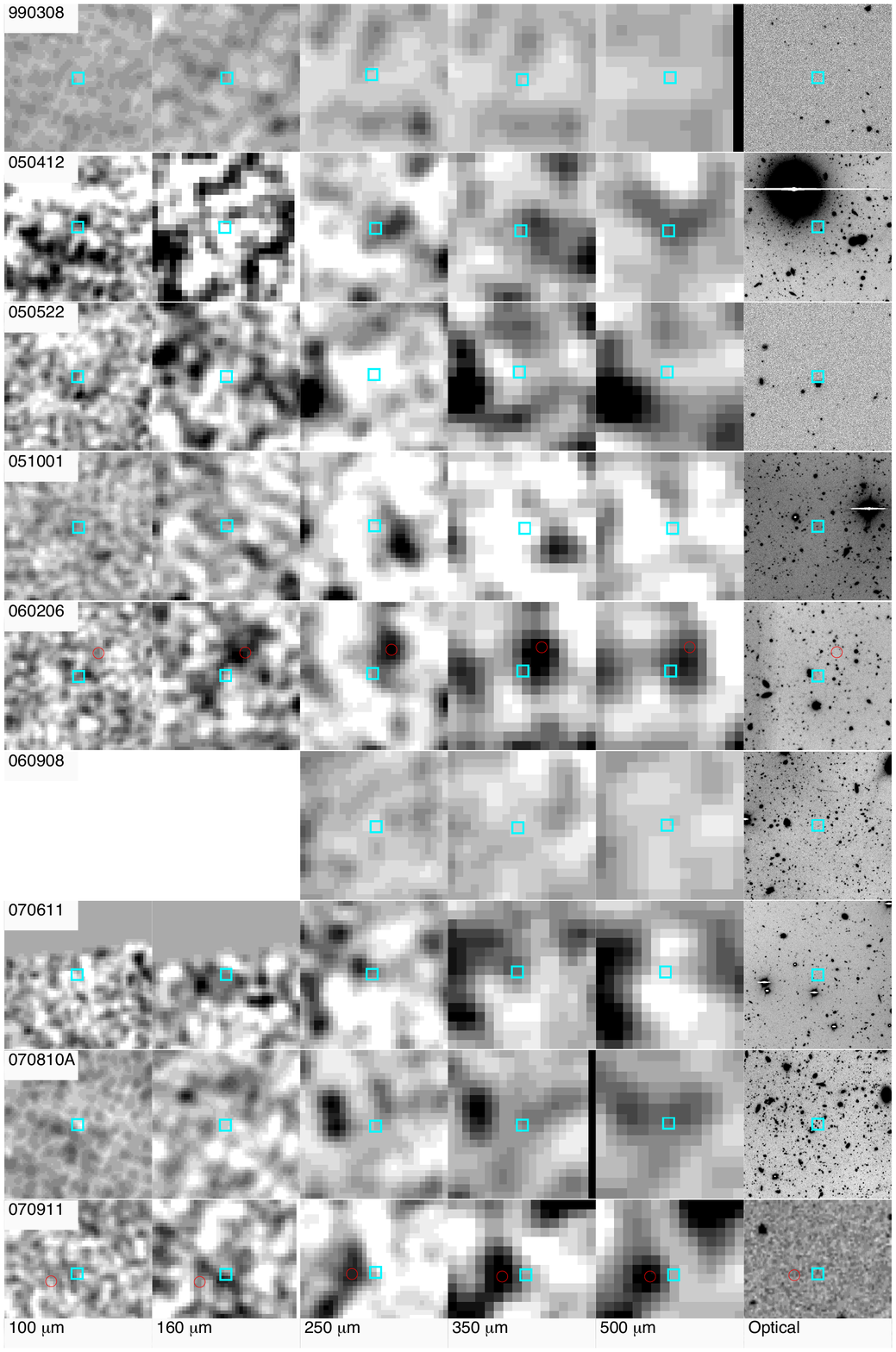}
\end{center}
\label{fig:samp}
\caption{{\it Herschel} images at the positions of the GRBs in our sample. At the centre of each frame, a blue box indicates the position of the GRB, as listed in Table~\ref{table_sample}. Additional sources, which were simultaneously fitted together with the GRB positions, are marked with red circles (see Section~\ref{res}). PACS images are shown from $-2$  (white) to $2$ mJy pixel$^{-1}$ (black). The pixel sizes are 3 and 4 arcsec for 100 and 160um, respectively. SPIRE images from $-10$ (white) to $30$ mJy beam$^{-1}$ (black). Each panel is 2 arcmin on a side. The first five columns show $100$, $160\,\mu$m PACS bands, and $250$, $350$, $500\,\mu$m SPIRE bands, respectively, while blank frames indicate that PACS maps of these regions (HerS) are not available. The last column shows the optical data (from \protect\cite{Hjorth.12} for 050412, 051001, 060908, 070611 and 070810A; from \protect\cite{Thone.08} for 060206; from \protect\cite{deUgarte.11} and \protect\cite{Sparre.11} for 110128A; from SDSS \protect\citep{Ahn.12} for 990308, 050522 and 110407A and from GAMA \protect\citep{Driver.09} for 070911, 080310, 091130B, 111204A, 120703A, 120927A, 121211A, 130502A, 140102A and 140515A).}
\end{figure*}
\addtocounter{figure}{-1}
\begin{figure*}
\begin{center}
\includegraphics[width=0.8\textwidth,clip]{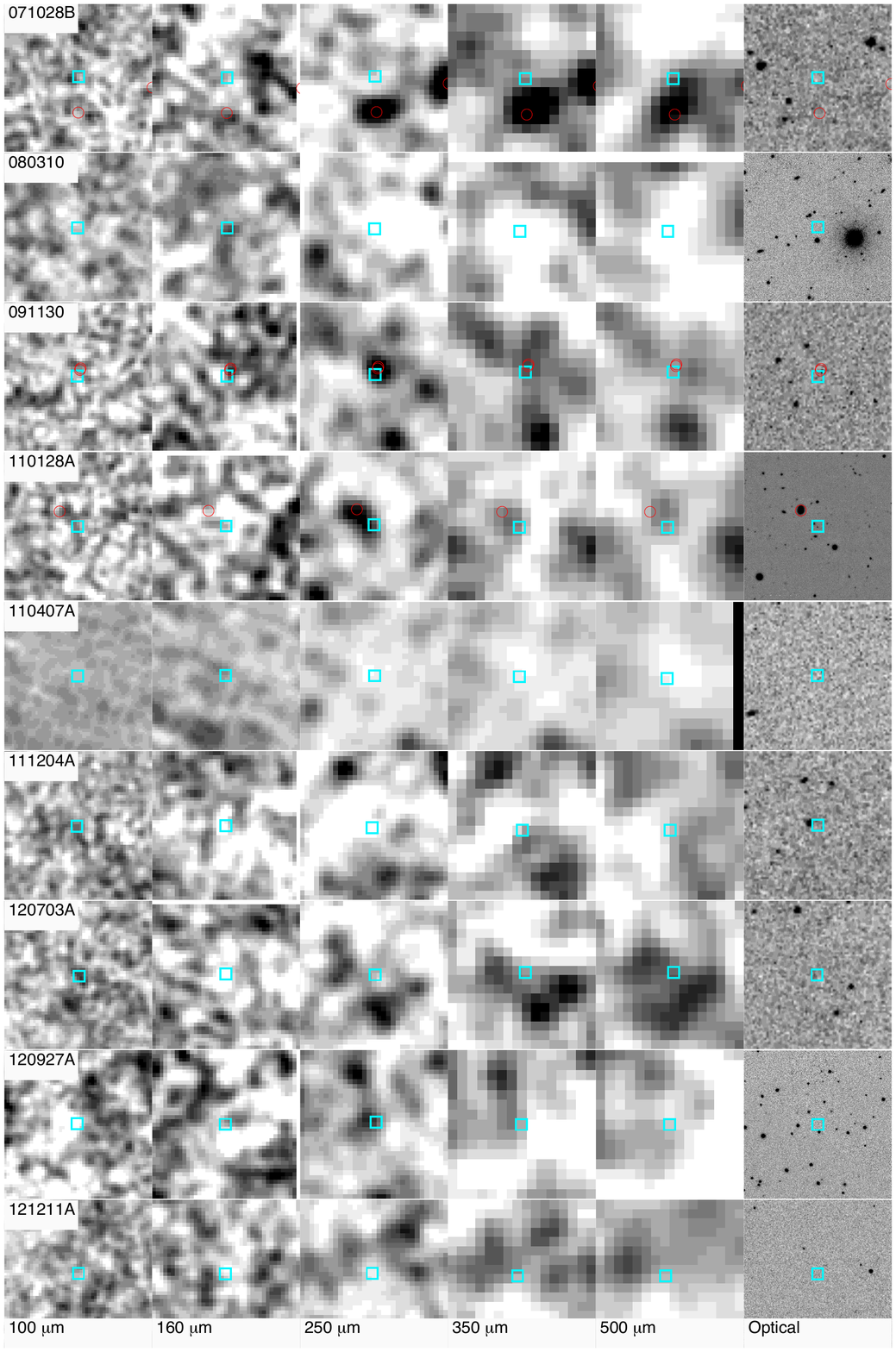}
\end{center}
\caption{(continued.)}
\end{figure*}
\addtocounter{figure}{-1}
\begin{figure*}
\begin{center}
\includegraphics[width=0.8\textwidth,clip]{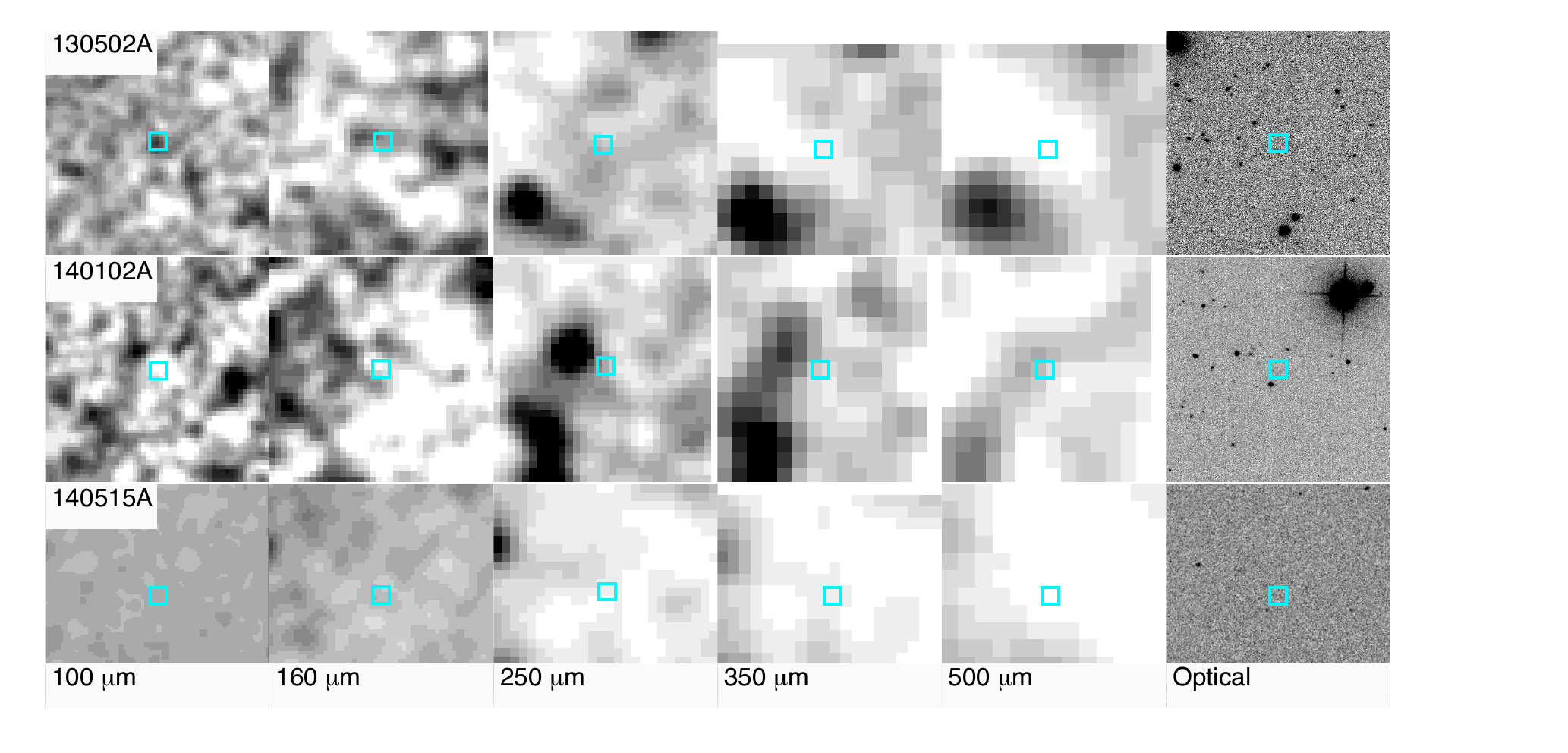}
\end{center}
\caption{(continued.)}
\end{figure*}

\section{Results}\label{res}

We determined the FIR flux densities of GRB hosts by fitting Gaussian functions at the GRB positions in the {\it Herschel} maps using the respective beam sizes\footnote{These were $9.4$, $13.4$, $18.2$, $24.9$ and $36.3$ arcsec for the $100$, $160$, $250$, $350$ and $500\,\mu$m bands, respectively.}. If a GRB host appeared to be detected at $>3\sigma$ we further investigated possible blending. We identified any obvious optical or FIR sources clearly distinct from the GRB host, and repeated the flux density measurement by fitting simultaneously two (or more) Gaussian functions at the GRB positions and at the positions of these other sources (this was performed for GRBs 060206, 070911, 071028B, 091130 and 110128A). For GRB 110128A in the deep optical image from \cite{deUgarte.11} and \cite{Sparre.11} we found a low-$z$ galaxy contributing to the {\it Herschel} flux density close to the GRB position (red circle on Fig.~1). No such obvious optical source was found close to the positions of GRB 060206 \citep{Thoene.07}, 071028B and 091130B \citep[GAMA;][]{Driver.09} but on the $250\,\mu$m images we identified sources that are clearly distinct from the GRB hosts. 

The flux densities in each band are presented in Table~\ref{table_flux}. Noise was estimated by measuring the flux at 100 random positions around the GRB position (and in the same way as for the GRB position) and the error is the $3\sigma$ clipped mean of these measurements; it was consistent with being dominated by the confusion noise \citep{Nguyen.10}.

Only GRB 091130B was located close to a source from the H-ATLAS catalogue ($9$ arcsec separation, corresponding to half of the beam size at $250\,\micron$). This source (HATLAS J133235.2+340526) has measured flux densities of $43\pm 6$, $21 \pm 7$ and $7 \pm 8\,$mJy in the $250$, $350$ and $500\,\mu$m bands, respectively \citep{Rigby.11}. An association between the GRB and the source is unlikely -- their  $9$ arcsec separation is large compared to the $2$ arcsec positional accuracy of the {\it Herschel} source.

 
SFRs were calculated according to \cite{Kennicutt.98}, based on the infrared luminosity integrated over $8\,-\,1000\,\mu$m using the SED of ULIRG Arp 220 \citep{Silva.98}, scaled to the flux densities of our GRB hosts. The use of Arp 220 is appropriate, since its dust temperature is close to that of typical GRB hosts \citep{Priddey.06, Michalowski.08, Michalowski.09.NotThesis, Michalowski.14, Watson.11, H.14, Hunt.14, Sym.14}. However, since we probe close to the peak of dust emission, the choice of the template does not influence the results substantially. For example, the median difference between the SFRs calculated using the SED of Arp 220 and the SED of a typical SMG \citep{Michalowski.10} is $\sim10$\%.

\begin{figure*}
\centering
\begin{tabular}{cc}
\includegraphics[scale=0.45]{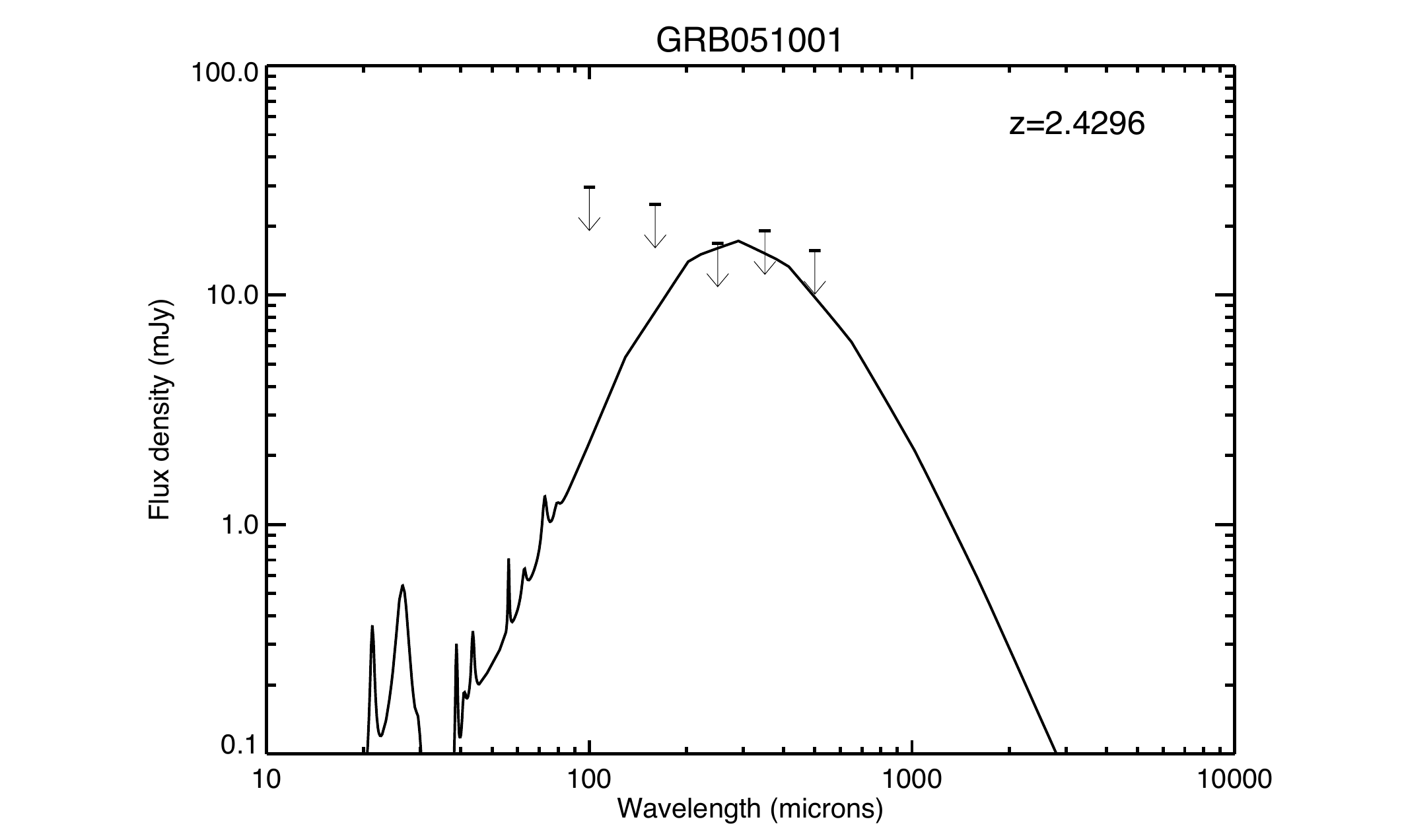}&\includegraphics[scale=0.45]{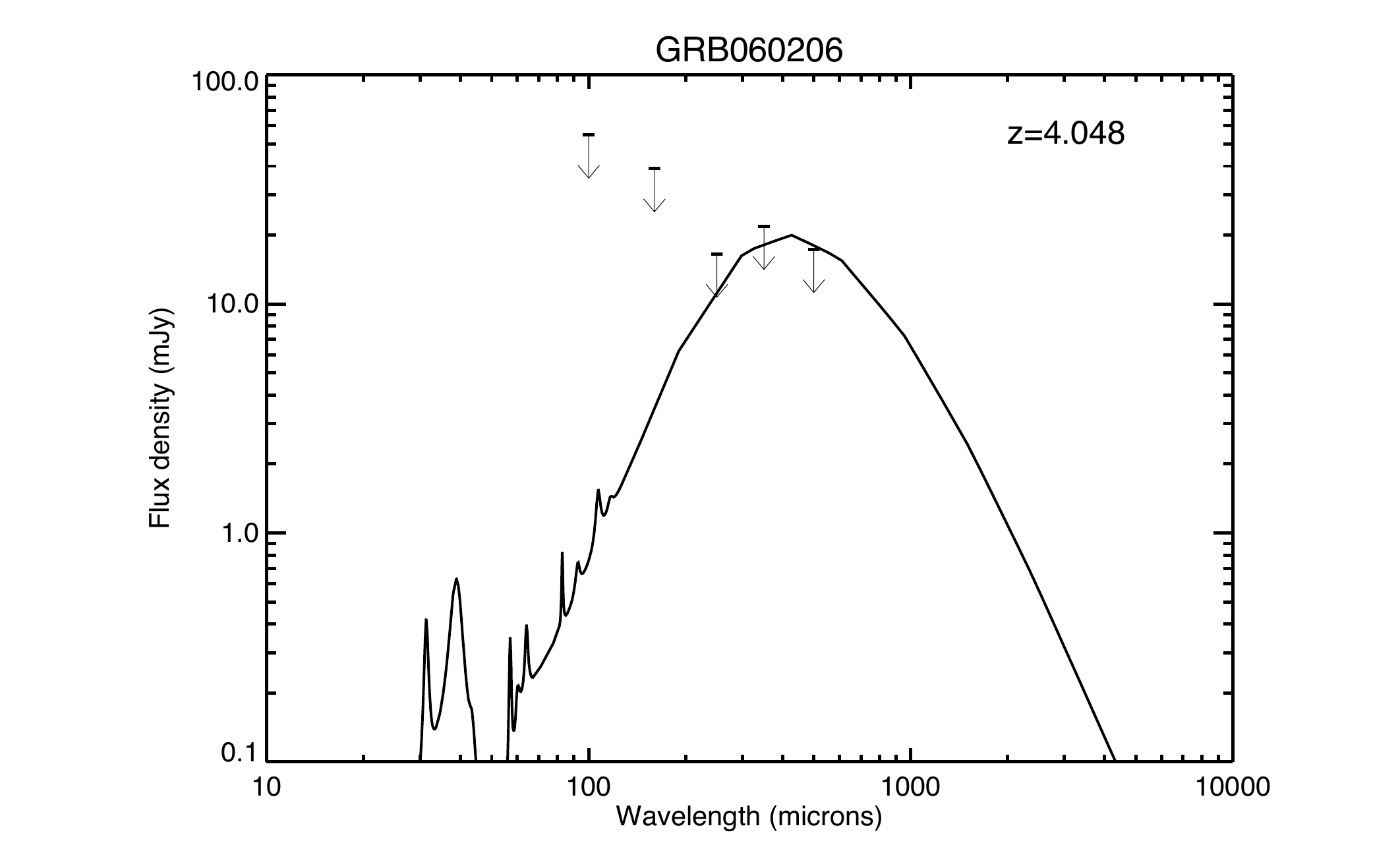} \\
\includegraphics[scale=0.45]{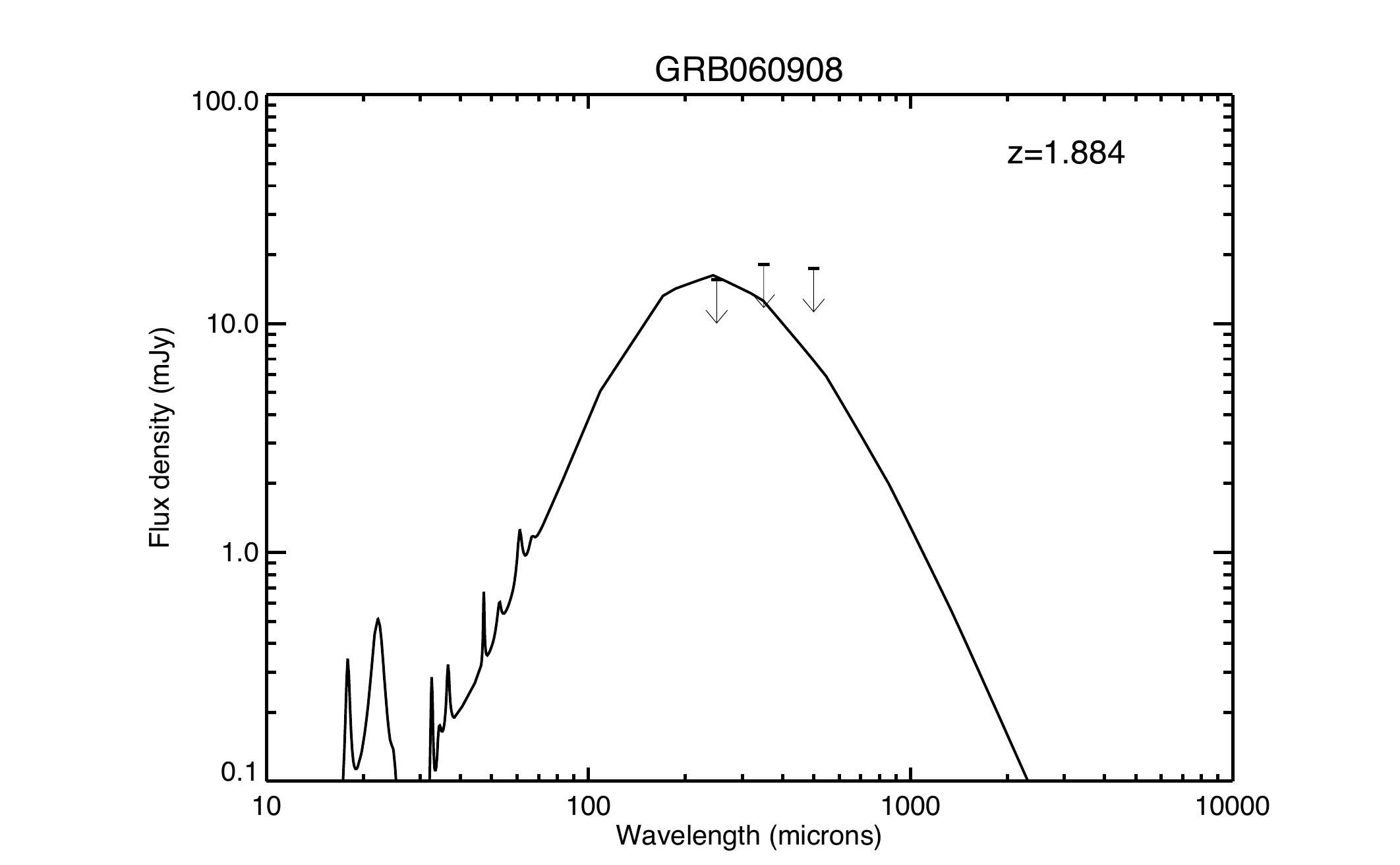}&\includegraphics[scale=0.45]{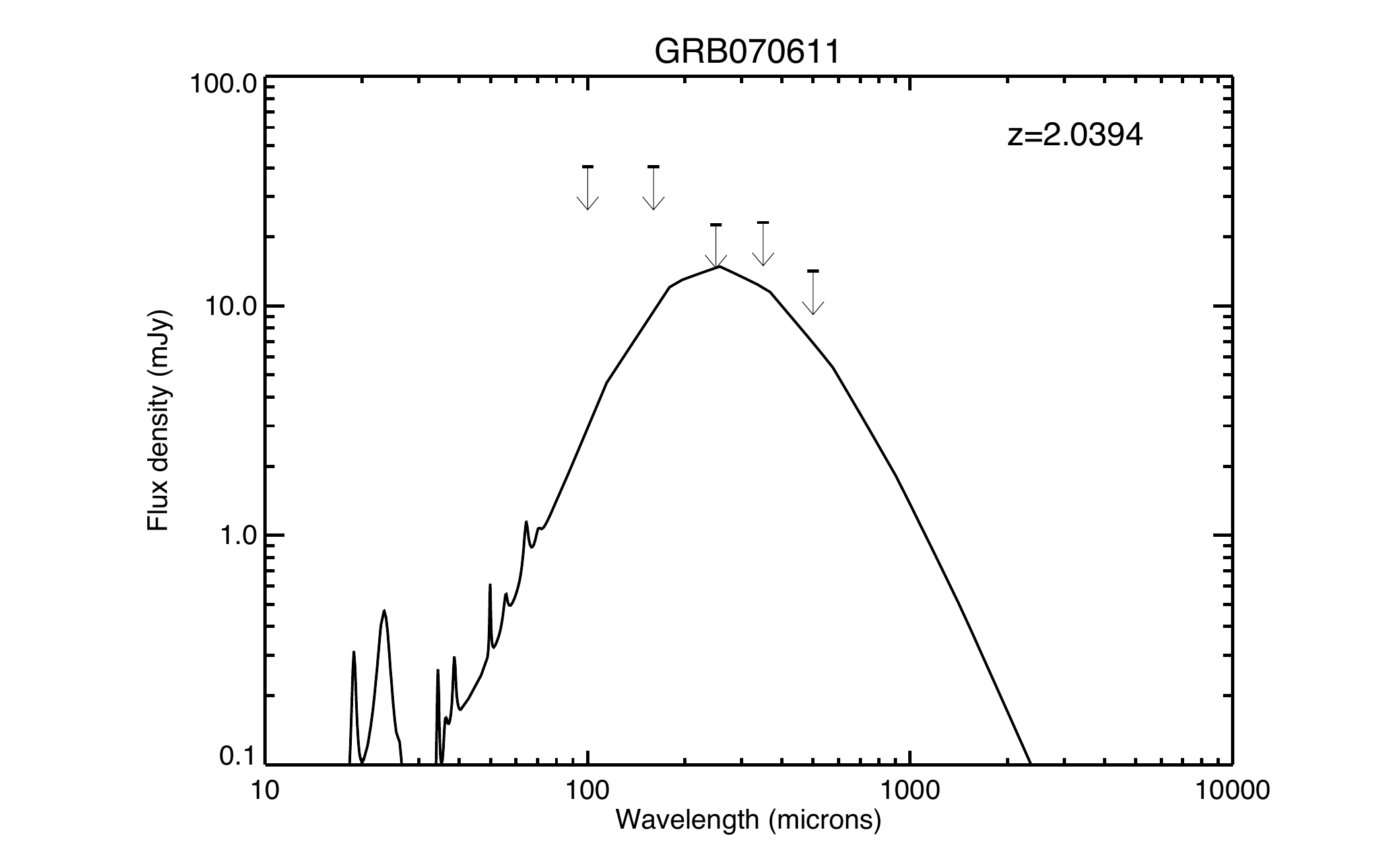} \\
\includegraphics[scale=0.45]{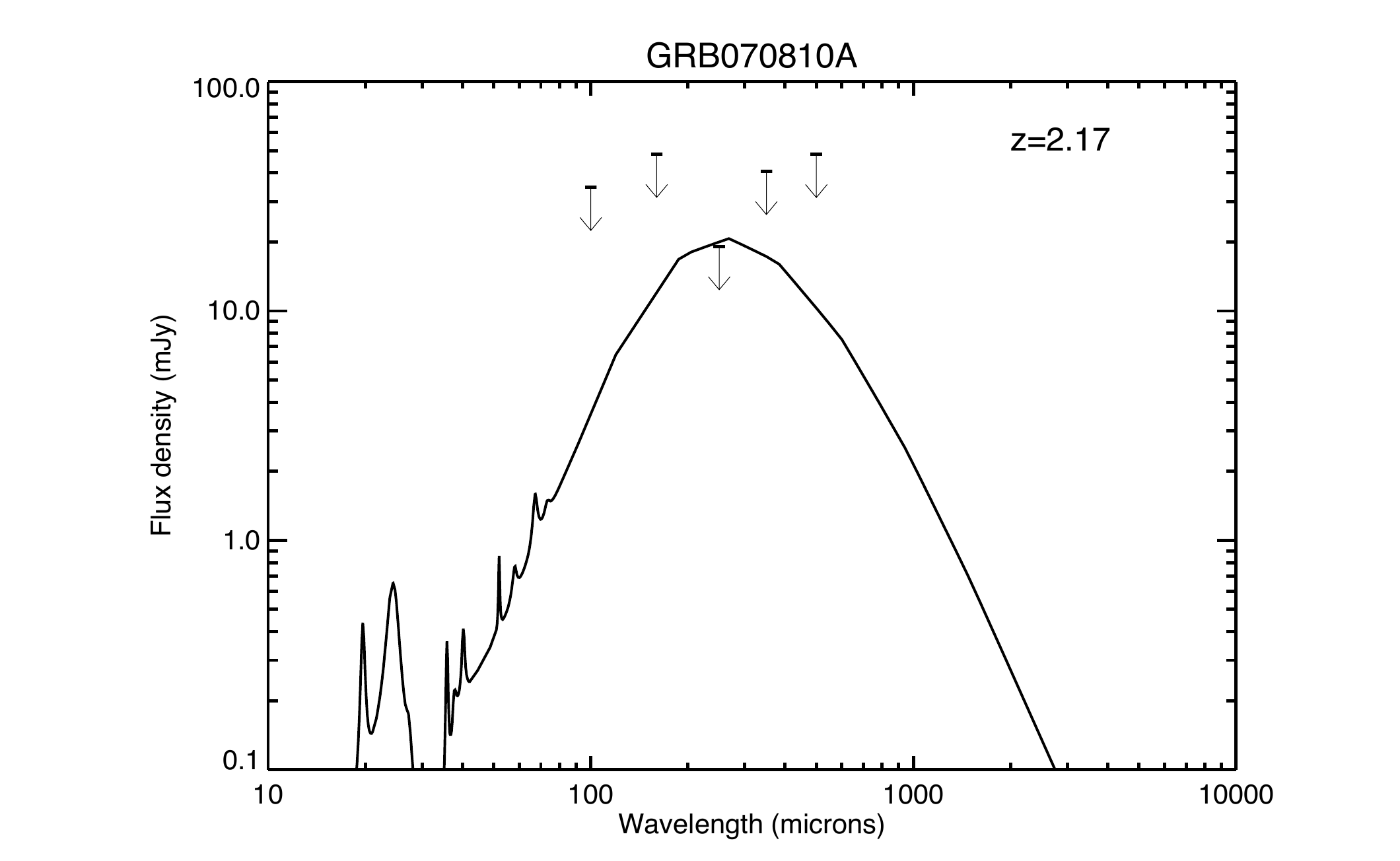}&\includegraphics[scale=0.45]{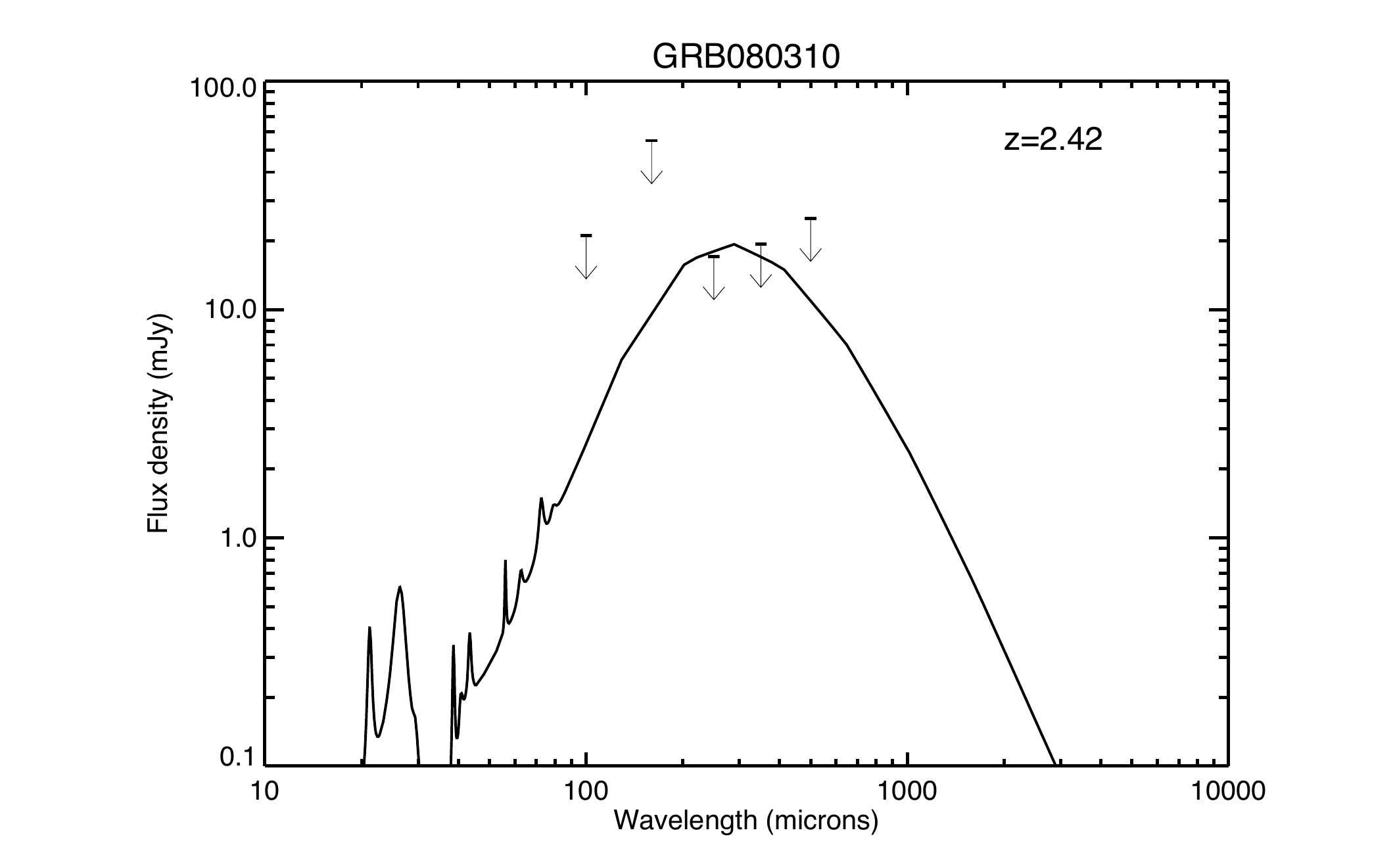}\\
\includegraphics[scale=0.45]{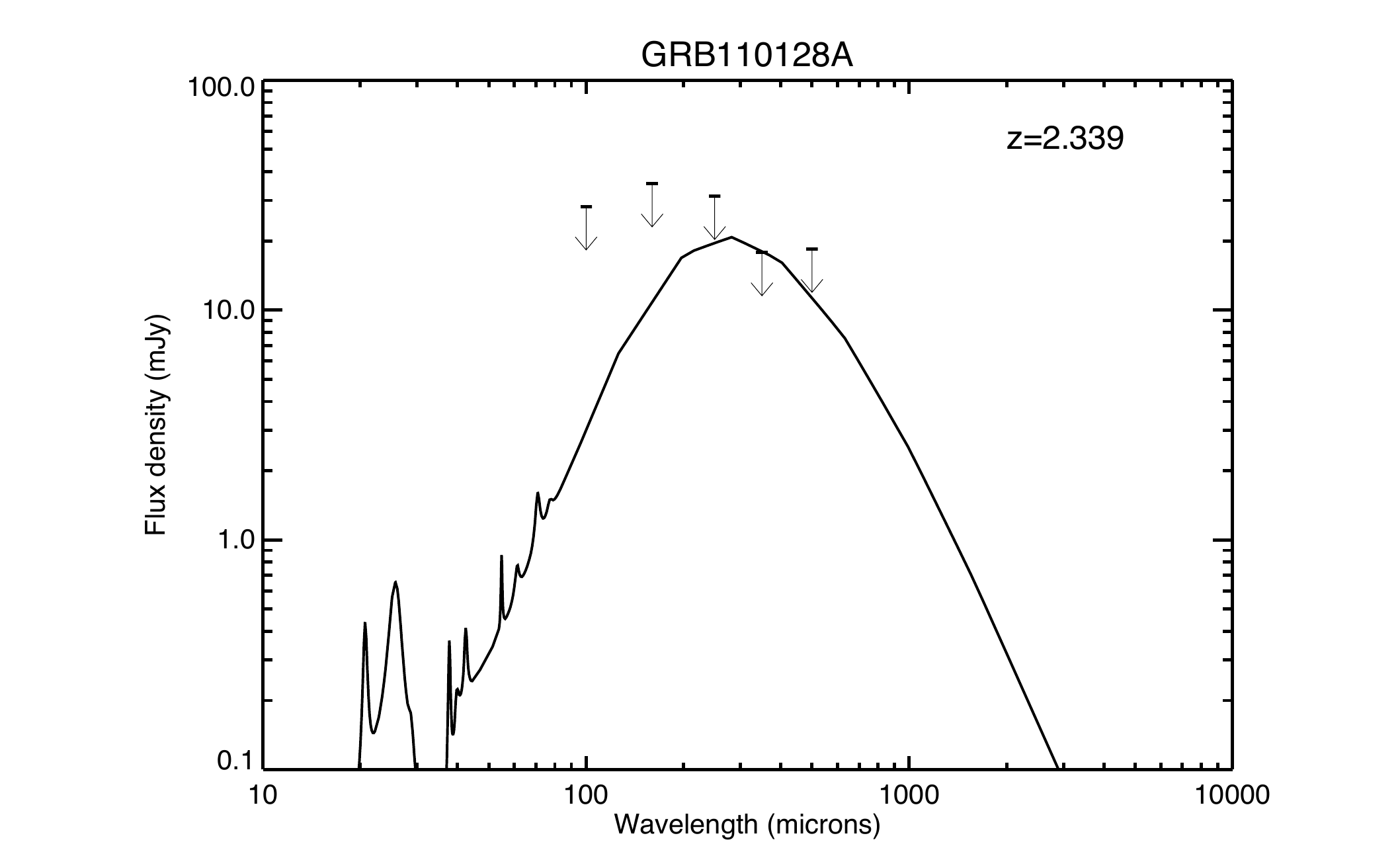}&\includegraphics[scale=0.45]{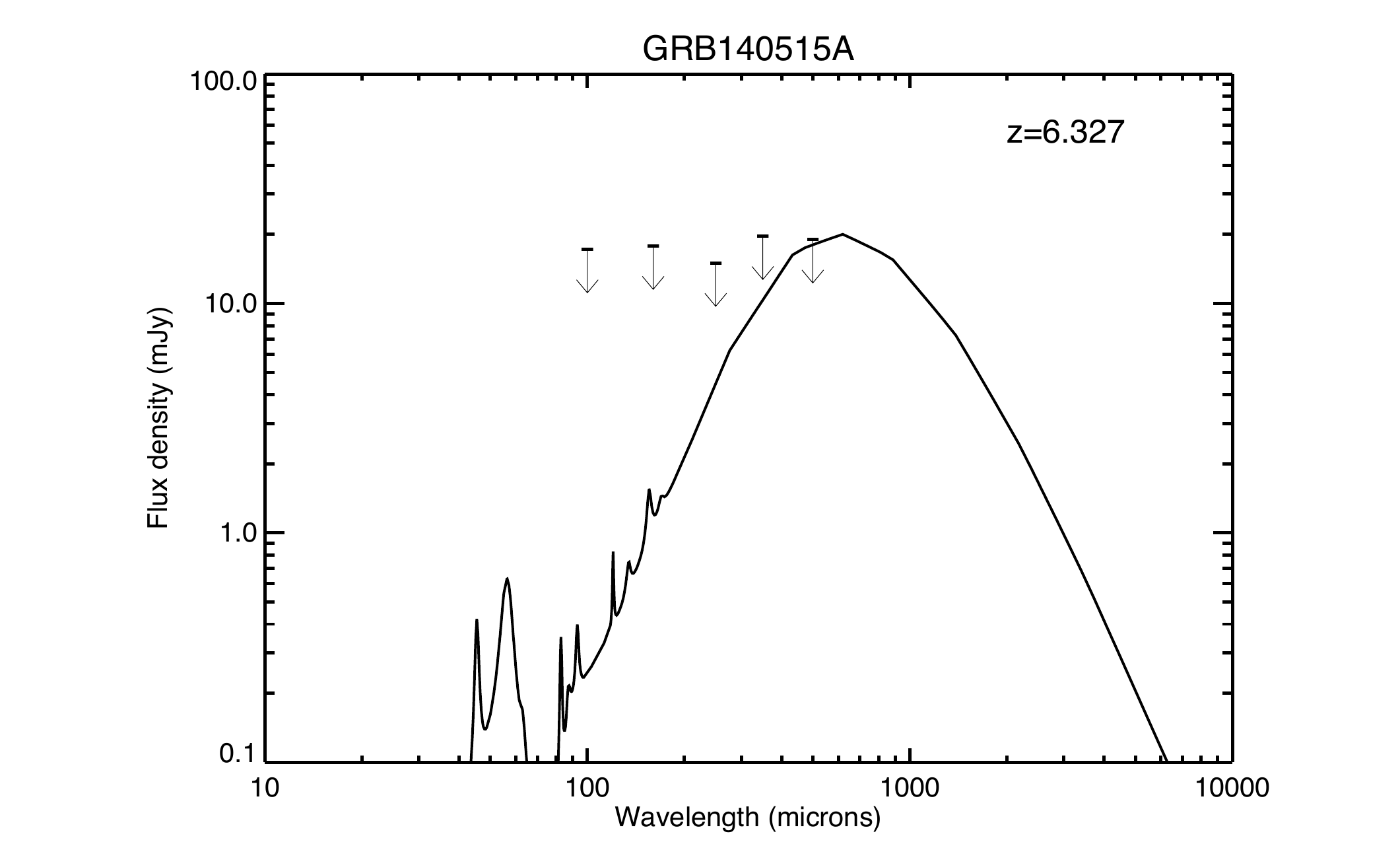} \\
\end{tabular}
\caption{$2\sigma$ upper limits on the SEDs of GRBs with known redshifts overlaid with the SED of Arp 220, used to calculate the SFR of each GRB host galaxy. The SED of Arp220 has been redshifted to match each GRB, shown on the upper right of each plot. Upper limits are from the base of the arrow.}
\label{fig_sed}
\end{figure*}

To derive dust masses of GRB hosts, we assumed $T_{\rm dust}=50\,$K \citep[a similar temperature to Arp 220;][]{Klaas.97, Dunne.00, Lisenfeld.00, Rangwala.11}, emissivity $\beta=1.5$ (which yields conservative upper limits) and a dust absorption coefficient $\kappa_{1.2\,{\rm mm}}=0.67\,\mbox{cm}^2\,\mbox{g}^{-1}$ \citep{Silva.98}. Assuming lower $T_{\rm dust}=30\,$K results in masses larger by a factor of $\sim10$. Derived properties of the GRB hosts (at $T_{\rm dust}=30$ and $50\,$K) with known redshift are presented in Table~\ref{table_zgals}.
Fig.~\ref{fig_sed} shows the spectral energy distributions of those GRBs with known redshifts, and are overlaid on the SED of Arp 220, which was used to calculate the SFR of each host.

\begin{table}
	\caption{{\it Herschel} flux density at GRB positions}
	\label{table_flux}
    \begin{tabular}{lccccc}
	\hline																					
	GRB	&	F100			&	F160			&	F250			&	F350			&	F500			\\
		&	(mJy)			&	(mJy)			&	(mJy)			&	(mJy)			&	(mJy)			\\
	\hline																					
	990308	&	$-$13	$\pm$	9	&	7	$\pm$	12	&	$-$1	$\pm$	7	&	2	$\pm$	8	&	$-$2	$\pm$	7	\\
	050412	&	$-$6	$\pm$	13	&	$-$24	$\pm$	18	&	7	$\pm$	7	&	2	$\pm$	8	&	9	$\pm$	7	\\
	050522	&	12	$\pm$	13	&	$-$17	$\pm$	17	&	$-$10	$\pm$	8	&	$-$7	$\pm$	8	&	0	$\pm$	9	\\
	051001	&	10	$\pm$	7	&	3	$\pm$	12	&	$-$5	$\pm$	8	&	$-$4	$\pm$	10	&	$-$6	$\pm$	8	\\
	060206	&	20	$\pm$	14	&	2	$\pm$	20	&	8	$\pm$	8	&	6	$\pm$	11	&	7	$\pm$	9	\\
	060908$^a$	&	--	&	--	&	$-$8	$\pm$	8	&	$-$3	$\pm$	9	&	$-$2	$\pm$	9	\\
	070611	&	$-$14	$\pm$	20	&	14	$\pm$	20	&	11	$\pm$	5	&	12	$\pm$	6	&	$-$3	$\pm$	7	\\
	070810A	&	$-$30	$\pm$	17	&	$-$19	$\pm$	24	&	5	$\pm$	10	&	15	$\pm$	10	&	28	$\pm$	12	\\
	070911	&	8	$\pm$	11	&	22	$\pm$	20	&	9	$\pm$	12	&	3	$\pm$	13	&	$-$7	$\pm$	11	\\
	071028B	&	6	$\pm$	14	&	0	$\pm$	20	&	$-$2	$\pm$	9	&	0	$\pm$	8	&	3	$\pm$	10	\\
	080310	&	$-$3	$\pm$	11	&	19	$\pm$	14	&	$-$12	$\pm$	9	&	$-$19	$\pm$	10	&	$-$24	$\pm$	13	\\
	091130	&	$-$9	$\pm$	14	&	$-$23	$\pm$	18	&	$-$11	$\pm$	9	&	4	$\pm$	8	&	4	$\pm$	8	\\
	110128A	&	13	$\pm$	14	&	$-$7	$\pm$	18	&	15	$\pm$	8	&	6	$\pm$	9	&	5	$\pm$	9	\\
	110407A	&	$-$10	$\pm$	11	&	11	$\pm$	18	&	$-$15	$\pm$	5	&	$-$14	$\pm$	9	&	$-$13	$\pm$	8	\\
	111204A	&	23	$\pm$	14	&	$-$34	$\pm$	17	&	$-$17	$\pm$	9	&	$-$3	$\pm$	13	&	5	$\pm$	11	\\
	120703A$^{b}$	&	33	$\pm$	13	&	-39	$\pm$	20	&	4	$\pm$	10	&	0	$\pm$	9	&	4	$\pm$	11	\\
	120927A	&	$-$20	$\pm$	13	&	0	$\pm$	22	&	16	$\pm$	11	&	$-$1	$\pm$	10	&	$-$9	$\pm$	12	\\
	121211A	&	$-$3	$\pm$	13	&	$-$16	$\pm$	16	&	$-$14	$\pm$	7	&	$-$3	$\pm$	8	&	$-$2	$\pm$	8	\\
	130502A	&	22	$\pm$	11	&	7	$\pm$	14	&	$-$10	$\pm$	5	&	$-$21	$\pm$	8	&	$-$31	$\pm$	9	\\
	140102A	&	$-$18	$\pm$	12	&	$-$13	$\pm$	17	&	$-$1	$\pm$	8	&	$-$6	$\pm$	9	&	$-$3	$\pm$	8	\\
	140515A	&	0	$\pm$	9	&	$-$8	$\pm$	9	&	$-$14	$\pm$	8	&	$-$13	$\pm$	10	&	$-$20	$\pm$	9	\\
	NWM$^{c}$ & 2 $\pm$ 3 & $-$4 $\pm$ 4 & $-$3 $\pm$ 2 & $-$2 $\pm$ 2 & $-$3 $\pm$ 2 \\
	Stacking & 1 $\pm$ 3 & $-$2 $\pm$ 4 & 3 $\pm$  2 & 3 $\pm$ 4 & 3 $\pm$ 2 \\
	\hline																					
    \end{tabular}
    \\
	Errors are from the clipped average of the flux densities of random positions around the GRB positions.\\
	$^a$ GRB 060908 is inside HerS, a SPIRE-only survey.\\
	$^{b}$ Tentative detection.\\
	$^{c}$ Noise-weighted mean.
\end{table}

Stacked images in the five {\it Herschel} bands were obtained by averaging
cut-out images around the central pixel corresponding to the
J2000 coordinates of each GRB host in the SPIRE and PACS images. Using
PSF-filtered images in each band, a region of $15\times15$ pixels was extracted
around each position, and these images were combined using a weighted
mean. Each pixel in the stacked image is the weighted mean of the
corresponding pixels from all the cut-outs, weighted by the inverse
variance of the flux density measurement at the GRB host position. The size of
the stacked images corresponds to $45, 60, 90, 120$ and $180$ arcsec at
each of the five bands, respectively. We did not detect any significant signal at any of the stacked
images (flux densities are presented in the last row of Table~\ref{table_flux}).

\section{Discussion}\label{disc}

GRBs are closely linked with star-formation, but they can be used as a probe of star formation only if the number of GRBs at a given redshift is proportional to the cosmic SFRD at this epoch. In such a case the distribution of SFRs of GRB hosts should reflect the contribution of galaxies to the SFRD. In particular, the fraction of hosts above some SFR threshold should be equal to the fractional contribution of such luminous galaxies to the SFRD. If the fraction of luminous GRB hosts is lower (higher) than this contribution, then we would conclude that GRBs are biased towards less (more) actively star-forming galaxies.

Integrating the $1.8<z<2.3$ infrared luminosity function of \citet{Magnelli.13} above $L_{\rm IR}>5\times10^{12}\,$L${\sun}$ ($\mbox{SFR} > 500\, {\rm M}_\odot\,\mbox{ yr}^{-1}$) we infer that such galaxies contribute $8^{+6}_{-3}$\% (a $2\sigma$ range of 3-24\%) to the cosmic SFRD at $z\sim2$. If GRBs trace star formation density in an unbiased way, we would expect this percentage of GRB hosts ($1.7_{-0.6}^{+1.3}$ hosts) in our sample to be detected, because {\it Herschel} sensitivity allows us to detect galaxies with L$_{\rm IR}> 5 \times 10^{12}{\rm L}_{\sun}$ (assuming the Arp 220 template at $z=2$--$2.5$ scaled to the $3\sigma$ limit of $15$--$20\,$mJy). 

Out of our 21 GRB hosts, we found just one tentative detection ($5\pm5$\%), namely GRB\,120703A with a significance of $\sim2.7\sigma$, suggesting that GRBs may trace star-formation density in an unbiased way, at least for redshift $2<z<4$ (and the expectation of $1-3$ detected hosts is an upper limit, since we are assuming that GRBs in our sample with no redshift information are at similar redshifts as the other ones). 
This is supporting evidence for the much sought-after relationship between GRB rate and SFR \citep[e.g.][]{Blain.00, Lloyd-Ronning.02, Hernquist.03, Christensen.04, Yonetoku.04, Yuksel.08, Kistler.09, Elliott.12, Michalowski.12, Robertson.12, Hunt.14}.

However, our analysis suffers from low number statistics as none of our {\it Herschel} flux measurements is of high significance, we consider here how our conclusions change if they are not real ($0^{+4}_{-0}$\% detection rate). That would produce a $<2\sigma$ tension between the data and the expectation that GRBs trace the star formation in an unbiased way at $z\sim2$ (i.e. $8^{+6}_{-3}$\% detection rate). This demonstrates that an FIR survey of a larger unbiased sample of GRB hosts is crucial to reach a statistically significant conclusion.

These conclusions are not uncontested. \cite{Perley.13} found a biased correspondence between the GRB host galaxy population and SFRD (at least at $z<1.5$). In their extensive study of the hosts of \textit{Swift}-observed GRBs, they found that GRB rate is a strong function of host-galaxy properties. In particular, GRB hosts were found to be less massive and bluer than what would be expected if the GRB rate is proportional to SFR in galaxies at $z<1.5$ \citep{Perley.13}. We probe GRB hosts at higher redshifts, but more significant samples of GRB hosts observed in the FIR are needed to investigate this issue.

The average 2$\sigma$ upper limit of the dust mass for our sample is log($M_{\rm dust}$/M$_{\sun}$)$=7.4 \pm 0.3\,$--$8.1 \pm 0.2$ ($50\,-30\,$K). This suggests that GRB hosts are less dusty than SMGs, as \cite{Michalowski.10} found the typical SMG to contain log($M_{\rm dust}$/M$_{\sun}$)$=9.02\pm0.36$ \citep[also see][]{Chapman.05, Kovacs.06, Laurent.06, Coppin.08, Magnelli.12, Swinbank.14}. The dust masses of GRB hosts in our sample are only consistent with those of SMGs if we assume a very low temperature of 30K (e.g. \citealt{Michalowski.10} found and average temperature of $35\,$K for SMGs). \cite{Calzetti.00} and \cite{Smith.13} found that dust emission at much lower temperatures ($20-30$K) could account for a large fraction of total flux in starburst galaxies, but other studies \citep{Priddey.06, Michalowski.08, Michalowski.09.NotThesis, Michalowski.14, Sym.14} suggest this is an unrealistically low temperature for an infrared-bright GRB host (however none of these studies is based on an unbiased GRB sample). Moreover, high dust temperature was found to be typical among {\it Herschel}-selected ULIRGs at high redshifts \citep{Sym.13}.

The noise-weighted average of the flux densities at all 20 GRB positions and flux densities measured in the stacked maps are shown in Table~\ref{table_flux}. 
At an average GRB redshift of $z=2.14$ \citep{Hjorth.14}, this corresponds to a 2$\sigma$ limit of SFR $< 114\, \mbox{M}_{\sun}\mbox{yr}^{-1}$ ($L_{\rm IR} < 2\times 10^{12}\,{\rm L}_\odot$) using the Arp 220 template, and 2$\sigma$ limits on dust masses of $M_{\rm dust}<2 \times10^{7}\mbox{M}_{\sun}$ for $T_{\rm dust}=50\,$K, and $M_{\rm dust}<2\times10^{8}\mbox{M}_{\sun}$ for $T_{\rm dust}=30\,$K. 
Measurements of stacked images provide consistent limits on the dust mass of typical GRB hosts: $M_{\rm dust}<6 \times10^{7}\mbox{M}_{\sun}$ for $T_{\rm dust}=50\,$K, and $M_{\rm dust}<2.4\times10^{8}\mbox{M}_{\sun}$ for $T_{\rm dust}=30\,$K. The deepest limit on SFR comes from the stacked $250\,\mu$m-band, giving a typical GRB host SFR of $<114$ M$_{\sun}$yr$^{-1}$.

If GRBs trace the cosmic SFRD in an unbiased way, then the average luminosity of GRB hosts should be equal to the average luminosity of other galaxies weighted by their SFRs (because a galaxy with a higher SFR will have a higher probability to host a GRB). From the luminosity function ($\phi$) of \cite{Magnelli.13} we calculated the weighted mean $\langle L\rangle_{\rm SFR}=\int_{L_{\rm min}}^{L_{\rm max}} \phi\cdot\mbox{SFR}\times L\,dL / \int_{L_{\rm min}}^{L_{\rm max}} \phi\times \mbox{SFR}\,dL$. Assuming $\mbox{SFR}\propto L_{\rm IR}$ this gives $\langle \log(L/{\rm L}_\odot)\rangle_{\rm SFR}=12.23\pm0.15$, or $\langle \mbox{SFR}\rangle_{\rm SFR}=290^{+120}_{-90}\,M_\odot\,\mbox{yr}^{-1}$ using the \cite{Kennicutt.98} conversion and assuming the Saltpeter IMF (propagating the errors on luminosity function parameters using the Monte Carlo method). This value is only weakly dependent on the adopted cut-off luminosities, $\log (L_{\rm min}/{\rm L}_{\odot})=7$ and $\log(L_{\rm max}/{\rm L}_{\odot})=13$, as it is mostly constrained by the shape of the luminosity function close to its knee. Again, this value is only $<2\sigma$ away from the limit we measured for GRB hosts ($<114\,{\rm M}_\odot\,\mbox{yr}^{-1}$), so we cannot rule out that GRB hosts are simply drawn based on the SFRs from the general population of galaxies, and we highlight that larger samples of unbiased GRB hosts observed at the FIR are needed.

The average flux density suggests that a typical GRB host has a flux below the confusion limit of {\it Herschel}, so surveys with better resolution are needed to detect the majority of the population. Such surveys will be undertaken in the future with the Large Millimeter Telescope and Cerro Chajnantor Atacama Telescope.

\begin{table*}
	\caption{Properties of GRBs with known redshifts}
	\label{table_zgals}
    \begin{tabular}{lllll}
    \hline
    GRB	& $z$ & Dust mass ($T=30$K) 	& Dust mass ($T=50$K) & SFR\\			
        &	& (10$^{8}$ M$_{\sun}$)& (10$^{8}$ M$_{\sun}$) & (M$_{\sun}$yr$^{-1}$)\\ 			
    \hline
	051001	&	2.4296 & $<$11 & $<$0.9 & $<$500 \\
	060206	&	4.048 & $<$24 & $<$2.7 & $<$1500 \\
	060908	&	1.884 & $<$6.2 & $<$0.7 & $<$320 \\
	070611	&	2.0394 & 5.1 $\pm$ 2.6 & 0.5 $\pm$ 0.2 & 280 $\pm$ 120 \\
	070810A	&	2.17 & $<$9.8 & $<$0.9 & $<$520 \\
	080310	&	2.42 & $<$11 & $<$1.0 & $<$560 \\
	110128A	&	2.339 & $<$9.4 & $<$0.9 & $<$550 \\
	140515A	&	6.327 & $<$83 & $<$4.9 & $<$2800 \\														
	NWM$^a$ &   2.14 & $<$1.8 & $<$0.3 & $<$114\\
	Stacking$^a$&   2.14 & $<$2.4 & $<$0.6 & $<$114\\
    \hline
    \end{tabular}
    \\
	Dust masses calculated assuming emissivity $\beta=1.5$. Errors reflect only statistical uncertainty propagated from the flux density errors, and limits are to $2\sigma$.\\
	$^a$ These are the properties calculated from the noise-weighted mean flux density of all 21 GRB positions, at an average GRB redshift of $z=2.14$ (\citealt{Hjorth.14}; Section~\ref{disc}).
\end{table*}
 
\section{Conclusion}\label{conc}

We have measured the FIR flux densities of $20$ GRB host galaxies. The sample was selected in a novel, unbiased fashion, in which we used a data base of GRB positions to cross-reference the positions surveyed by the H-ATLAS, HeViCS, HeFoCS, HerMES and HerS data releases. For 8 out of the 20 GRBs, redshifts are available, and for these we were able to calculate or place an upper limit on the SFR and dust mass of the host galaxy. One host was (tentatively) detected, consistent with the contribution of such bright galaxies to the SFRD in the Universe at $z\sim2$ \citep{Magnelli.11, Magnelli.13}. This may support the thesis that the GRB rate and SFR are fundamentally related, and that GRBs may trace SFRD in an unbiased way. Analysing a larger sample of GRB hosts selected in such an unbiased way is necessary to give a definite answer on whether this is indeed the case.

\section*{Acknowledgements}\label{ackno}

We thank Daniele Malesani and Christina Th{\"o}ne for their contributions of additional optical data, and the anonymous referee for their helpful comments, which improved the quality of the paper. MJM acknowledges the support of the UK Science and Technology Facilities Council. NB is supported by the EC FP7 SPACE project ASTRODEEP (Ref. no. 312725). EI acknowledges funding from CONICYT/FONDECYT postdoctoral project N$^\circ$:3130504. LD, RJI and SJM acknowledge support from ERC Advanced Grant COSMICISM. JGN acknowledges financial support from the Spanish CSIC for a JAE-DOC fellowship, co-funded by the European Social Fund, by the Spanish Ministerio de Ciencia e Innovacion, AYA2012-39475-C02-01, and Consolider-Ingenio 2010, CSD2010-00064, projects.

The {\it Herschel}-ATLAS is a project with {\it Herschel}, which is an \textit{ESA space observatory} with science instruments provided by European-led Principal Investigator consortia and with important participation from NASA. The H-ATLAS website is \url{http://www.h-atlas.org/}.

GAMA is a joint European-Australasian project based around a spectroscopic campaign using the Anglo-Australian Telescope. The GAMA input catalogue is based on data taken from the SDSS and the UKIRT Infrared Deep Sky Survey. Complementary imaging of the GAMA regions is being obtained by a number of independent survey programmes including \textit{GALEX} MIS, VST KIDS, VISTA, VIKING, \textit{WISE}, {\it Herschel}-ATLAS, GMRT and ASKAP providing UV to radio coverage. GAMA is funded by the STFC (UK), the ARC (Australia), the AAO, and the participating institutions. The GAMA website is: \url{http://www.gama-survey.org/}.

This research has made use of data from HerMES project (\url{http://hermes.sussex.ac.uk/}). HerMES is a {\it Herschel} Key Programme utilizing Guaranteed Time from the SPIRE instrument team, ESAC scientists and a mission scientist. The HerMES data were accessed through the {\it Herschel} Database in Marseille (HeDaM - \url{http://hedam.lam.fr}) operated by CeSAM and hosted by the Laboratoire d'Astrophysique de Marseille.

Funding for SDSS-III has been provided by the Alfred P. Sloan Foundation, the Participating Institutions, the National Science Foundation, and the US Department of Energy Office of Science. The SDSS-III web site is \url{http://www.sdss3.org/}.
\bibliographystyle{mn2e}
\bibliography{grbbib}{}

\label{lastpage}

\end{document}